\documentclass[12pt]{article}
\usepackage{graphicx}
\usepackage{amsfonts}
\usepackage{latexsym}
\title{Modal quantum theory}
\author{Benjamin Schumacher\footnote{Department of Physics, Kenyon College.  Email schumacherb@kenyon.edu} 
    \,\, and Michael D. Westmoreland\footnote{Department of Mathematical Sciences, Denison University.  Email westmoreland@denison.edu}}
\date{Kenyon College and Denison University}
\newcommand{\mket}[1]{\left | #1 \right ) }
\newcommand{\mbra}[1]{\left ( #1 \right | }
\newcommand{\mamp}[2]{\left ( #1 \left | #2 \right. \right ) }

\newcommand{\modalspace}{\mbox{$\mathcal{V}$}}

\newcommand{\scalarfield}{\mbox{$\mathcal{F}$}}
\newcommand{\zeetwo}{\mbox{$\mathbb{Z}_{2}$}}

\newcommand{\possible}[2]{\mathcal{P} \! \left ( #1 | #2 \right )}
\newcommand{\hvpossible}[2]{\mathcal{P}_{#2} \! \left ( #1 \right )}
\newcommand{\vspan}[1]{\left \langle #1 \right \rangle}
\newcommand{\dual}[1]{{#1}^{\ast}}

\newcommand{\ket}[1]{\left | #1 \right \rangle}

\newcommand{\rootonehalf}{\frac{\scriptstyle 1}{\scriptstyle \sqrt{2}}}

\newcommand{\sys}[1]{_{\mbox{\tiny #1}}}

\begin{document}
\maketitle
\thispagestyle{empty}

\begin{abstract}
  We present a discrete model theory similar in structure 
  to ordinary quantum mechanics, but based on a finite field 
  instead of complex amplitudes.
  The interpretation of this theory involves only the ``modal''
  concepts of possibility and necessity rather than quantitative
  probability measures.  
  Despite its simplicity, our model theory includes entangled 
  states and has versions of both Bell's theorem and the 
  no cloning theorem.
\end{abstract}

\section*{Modal quantum theory}

In quantum theory, the states of physical systems are represented
by vectors in a complex Hilbert space.  The complex scalars serve
as probability amplitudes, quantities whose squared magnitudes
are the probabilities of measurement outcomes.  Other types
of quantum theory have sometimes been considered, based on
real or quaternionic amplitudes \cite{finkelstein,aaronson}.
Though the quantum
mechanics of nature does not appear to be real or quaterionic,
these alternate mathematical formalisms shed light on the
structure of the actual quantum theory (which we will here
abbreviate AQT).

Here we will explore the properties of another
type of ``toy model'' of quantum theory using scalars
drawn from a finite field \scalarfield.  The simplest example
is based on the two-element field $\zeetwo$, but many
other choices are possible.  Our toy model lacks much
of the mathematical paraphernalia of complex Hilbert
spaces.  For instance, there is no natural inner
product and thus no concept of ``orthogonality'' between
vectors.  Nevertheless, we will find that the theory is
well-defined, that it has a sensible interpretational
framework, and that entanglement and many other aspects
of AQT have analogues in the theory.

The interpretation of AQT involves quantitative probabilities,
but our interpretation of finite-field theories is more
primitive, involving only the distinction between
{\em possible} and {\em impossible} events.  Suppose
$\mathcal{E}$ is the set of outcomes of some experiment.
In AQT, a given quantum state would yield a probability
distribution over the elements of $\mathcal{E}$.  But our
new theory will only designate a non-empty subset
$\mathcal{P} \subseteq \mathcal{E}$, the set of possible results,
without distinguishing more or less likely elements of the
set.  Any outcome not in $\mathcal{P}$ is taken to be
impossible, and if $\mathcal{P}$ only contains a single
element $r$, then we may say that $r$ is ``certain''
or ``necessary''.

This distinction between ``possible'', ``impossible''
and ``necessary'' events is exactly the distinction
used in modal logic \cite{modallogic}.  Thus, we will
refer to our finite-field quantum theories as {\em modal
quantum theory}, or MQT.

For a finite field \scalarfield, the MQT state of a system
is a non-null vector $\mket{\psi}$ in a finite-dimensional
vector space $\modalspace$, which is isomorphic to $\scalarfield^{d}$
for some dimension $d$.  A measurement on the system corresponds
to a basis set $A = \{ \mket{a} \}$ for $\modalspace$, where each basis
element $\mket{a}$ is associated with an outcome $a$ of the
measurement procedure.  (Note that, in the absence of an inner
product, there is no requirement in MQT that the basis elements
be orthogonal.)  Every state vector $\mket{\psi}$ can be
written
\begin{equation}
    \mket{\psi} = \sum_{a} \psi_{a} \mket{a} ,  \label{eq:expandinbasis}
\end{equation}
where the coefficients $\psi_{a}$ are scalars in
$\scalarfield$.  The measurement outcome $a$ is possible
if and only if $\psi_{a} \neq 0$.  For the basis $A$ and
state $\mket{\psi}$, the set of possible measurement
results is thus
\begin{equation}
    \possible{A}{\psi} = \{ a : \psi_{a} \neq 0 \} .
\end{equation}

The simplest type of MQT has $\scalarfield = \zeetwo$,
and the simplest MQT system has state space dimension $d = 2$.
The resulting example may be called a {\em mobit}.
A mobit has three states:  basis states $\mket{0}$ and $\mket{1}$,
and a single superposition state $\mket{\sigma} = \mket{0} + \mket{1}$.
In fact, any one of these states is a superposition of the other
two, and so any pair of the states is a basis for the vector
space.  We define three modal observables, which
we will call $X$, $Y$ and $Z$, associated with the
three possible basis sets.  For each measurement, we
can conveniently label the two outcomes by $+$ and $-$.
That is,
\begin{equation}
    \begin{array}{ccccc}
        \mket{+_{z}} = \mket{0} & \quad &
        \mket{+_{x}} = \mket{1} & \quad &
        \mket{+_{y}} = \mket{\sigma} \\
        \mket{-_{z}} = \mket{1} & \quad &
        \mket{-_{x}} = \mket{\sigma} & \quad &
        \mket{-_{y}} = \mket{0}
    \end{array} .
    \label{xyzbases}
\end{equation}
The question of whether a basis element corresponds to a 
possible outcome generally depends on the entire basis.
For example, consider the mobit state $\mket{\sigma}$.
If we measure $Z$ then the result $(+_{z})$ corresponding
to the basis vector $\mket{0}$ is possible.  However,
if we measure $Y$ then the result $(-_{y})$ corresponding
to the same basis vector $\mket{0}$ is not possible.

Things are clearer if we associate a measurement with a
basis of the dual space $\dual{\modalspace}$.  Every basis
$\{ \mket{a} \}$ for $\modalspace$ is associated with a dual
basis $\{ \mbra{a} \}$ for $\dual{\modalspace}$ such that
$\mamp{a}{\psi} = \psi_{a}$, the component of $\mket{\psi}$
in Equation~\ref{eq:expandinbasis}.  (In the absence of
an inner product, the correspondence between $\mket{a}$
and $\mbra{a}$ is basis-dependent.)  We call the functionals
in $\dual{\modalspace}$ {\em effects} and say
that an effect $\mbra{a}$ is possible given the
state $\mket{\psi}$ provided $\mamp{a}{\psi} \neq 0$.
Thus, given a basis $A$ for $\dual{\modalspace}$,
\begin{equation}
    \possible{A}{\psi} = \{ a : \mamp{a}{\psi} \neq 0 \} .
\end{equation}
The question of whether a particular effect is possible
does not depend on the dual basis to which it belongs.

We can express the $X$, $Y$ and $Z$ mobit measurements using 
dual bases.  Let $\{ \mbra{0}, \mbra{1} \}$ be the dual basis 
corresponding to the $\{ \mket{0}, \mket{1} \}$ basis.
Thus $\mamp{0}{0} = \mamp{1}{1} = 1$ and $\mamp{0}{1} =
\mamp{1}{0} = 0$.  The remaining dual vector is
$\mbra{\sigma} = \mbra{0} + \mbra{1}$.  Then we have
\begin{equation}
    \begin{array}{ccccc}
        \mbra{+_{z}} = \mbra{0} & \quad &
        \mbra{+_{x}} = \mbra{\sigma} & \quad &
        \mbra{+_{y}} = \mbra{1} \\
        \mbra{-_{z}} = \mbra{1} & \quad &
        \mbra{-_{x}} = \mbra{0} & \quad &
        \mbra{-_{y}} = \mbra{\sigma}
    \end{array} .
    \label{xyzdualbases}
\end{equation}
Compare Equation~\ref{xyzbases}.

Finally, we can outline a framework for describing the
time evolution of a system in MQT.  Time must be regarded
as a sequence of discrete intervals.  Just as in AQT, the
``coherent'' time evolution of a system over one of these
intervals is represented by a linear transformation $T$ of the
state vector.  Thus, if $\mket{a} \rightarrow \mket{a'} = T \mket{a}$
and $\mket{b} \rightarrow \mket{b'} = T \mket{b}$ then
\begin{equation}
    \mket{a} + \mket{b} \rightarrow T \left ( \mket{a} + \mket{b} \right )
        = T \mket{a} + T \mket{b} = \mket{a'} + \mket{b'} .
\end{equation}
Since the zero vector is not a physical state, we require that
$T \mket{a} \neq 0$ for any state $\mket{a}$.  This means that
the kernel of $T$ is trivial, so that $T$ is invertible.

No additional restriction (such as unitarity in AQT) on the time
evolution operator $T$ is motivated by the general framework
of MQT.  We will generally suppose that any invertible
linear transformation of state vectors corresponds to a possible
time evolution of the system.

\section*{Entangled states}

In AQT, the Hilbert space describing a composite system
is the tensor product of the Hilbert spaces describing
the individual subsystems.  The same rule applies to the
vector spaces in MQT.  In general, a composite system may
have both product states and non-product (entangled) states.
Since the state spaces in MQT are discrete, we can calculate
the numbers of product and entangled states for a given pair
of systems.  We find that every composite system has both
product and entangled states, and that as the subsystem
state space dimensions become large, the entangled states
greatly outnumber the product states.

Consider a pair of mobits, for which $\scalarfield = \zeetwo$.
There are 15 allowed state vectors for the pair, all representing
distinct states of the system.  Nine of these are product
states and six are entangled.

One particular entangled state of two mobits has especially
elegant properties:  $\mket{S} = \mket{0,1} + \mket{1,0}$.
Any product effect $\mbra{a,a}$ is impossible for $\mket{S}$
because
\begin{equation}
    \mamp{a,a}{S}
    = \mamp{a}{0} \mamp{a}{1} + \mamp{a}{1} \mamp{a}{0}
    = 0
\end{equation}
(recalling that $x + x = 0$ in $\zeetwo$).
From the dual basis forms of the $X$, $Y$ and $Z$ measurements
given in Equation~\ref{xyzdualbases}, we can draw the following
conclusions:
\begin{itemize}
\item  If the same measurement is made on each mobit, then
    the only possible joint results have opposite values 
    for each mobit.  For example, $(+_{z},-_{z})$ is possible;
    the result $(+_{z},+_{z})$ is impossible, since
    $\mbra{+_{z},+_{z}} = \mbra{0,0}$.
\item  If different measurements are made on the mobits,
    then there is one joint result that is impossible.
    For example, $(+_{z},-_{x})$ is impossible, since
    $\mbra{+_{z},-_{x}} = \mbra{0,0}$.
\end{itemize}
Since corresponding measurements must lead to opposite
results, the state $\mket{S}$ is analogous to the
``singlet'' state
$\rootonehalf \left ( \ket{\uparrow \downarrow} - \ket{\downarrow \uparrow} \right )$
of a pair of spins in AQT.

In AQT, Bell showed that the correlations between entangled quantum
systems were incompatible with any local hidden variable theory \cite{bell}.
He did this by devising a statistical inequality that must hold for
local hidden variable theories but is violated by entangled
quantum systems.  Is there an analogous result for MQT?
Unfortunately, in the absence of probabilities and expectation
values the Bell approach will not work.  However, Hardy \cite{hardy}
devised an alternate approach based only on possibility and
impossibility.

Hardy constructs a non-maximally entangled state $\ket{\Psi}$
of a pair of qubits, together with binary observables $A$ and $B$
on each qubit.  If we denote by $(x,y|X,Y)$ the outcome $(x,y)$
of a joint measurement $(X,Y)$, then Hardy's state has the
following properties.
\begin{itemize}
\item  $(0,0|A,B)$ and $(0,0|B,A)$ are both impossible---that is,
    they have quantum probability $p = 0$.
\item  $(0,0|B,B)$ is possible ($p > 0$).
\item  $(1,1|A,A)$ is impossible ($p = 0$).
\end{itemize}
How might a local hidden variable theory account for this situation?
Since $(0,0|B,B)$ is possible, we may restrict our attention to the
set $H$ of values of the hidden variables that lead to this result.
The result of a measurement on one qubit is unaffected by a change 
in the choice of measurement on the other (locality).  Furthermore,
no allowed values of the hidden variables can lead to $(0,0|A,B)$ 
or $(0,0|B,A)$.  Thus, for values in $H$, we would have to obtain
the results $(1,0|A,B)$ and $(0,1|B,A)$.  But these jointly imply
that the result $(1,1|A,A)$ would be obtained for values in $H$,
so that this result must be possible.  This contradicts AQT.

In the same way, we can show that the structure of possible and
impossible measurement results arising from the entangled mobit
state $\mket{S}$ above is incompatible with any local hidden
variable theory.

In a hidden variable theory, we imagine that the MQT state $\mket{S}$
corresponds to a set $H$ of possible values of a hidden variable.
We further imagine that the hidden variable controls the outcomes
of the possible measurements on the mobits in a completely local way.
That is, for any particular value $h \in H$,
the set of possible results of Alice's measurement depends
only on $h$ and her own choice of measurement,
not any measurement choices or results for Bob's mobit.
Let $\hvpossible{E}{h}$ be the set
of possible results of a measurement of $E$ for the hidden variable
value $h$.  Our locality assumption means that, given $V\sys{A}$ and
$W\sys{B}$ measurements for Alice and Bob and a particular $h$ value,
\begin{equation}
    \hvpossible{V\sys{A},W\sys{B}}{h} = \hvpossible{V\sys{A}}{h} \times \hvpossible{W\sys{B}}{h},
\end{equation}
the simple Cartesian product of separate sets $\hvpossible{V\sys{A}}{h}$ and
$\hvpossible{W\sys{B}}{h}$.The MQT set of possible results arising from
$\mket{S}$ should therefore be
\begin{equation}
    \possible{V\sys{A},W\sys{B}}{S} = \bigcup_{h \in H}
            \hvpossible{V\sys{A}}{h} \times \hvpossible{W\sys{A}}{h} .
\end{equation}
The individual sets $\hvpossible{V\sys{A}}{h}$, etc., are simultaneously 
defined for all of the measurements that can be made by Alice and Bob.  
Therefore, we may consider the set
\begin{eqnarray}
    \mathcal{J} & = &  \bigcup_{h \in H} \,\,
        \hvpossible{X\sys{A}}{h} \times \hvpossible{Y\sys{A}}{h}
        \times \hvpossible{Z\sys{A}}{h} \nonumber \\
        &  &  \qquad {} \times \hvpossible{X\sys{B}}{h}
        \times \hvpossible{Y\sys{B}}{h} \times \hvpossible{Z\sys{B}}{h} .
\end{eqnarray}
There might be up to $2^{6} = 64$ elements in $\mathcal{J}$.  However,
since $\mathcal{J}$ can only contain elements that agree with the
properties of $\mket{S}$, we can eliminate many elements.  For instance,
the fact that corresponding measurements on the two mobits must give
opposite results tells us that $(+,+,+,+,+,+)$ cannot be in $\mathcal{J}$,
though $(+,+,+,-,-,-)$ might be.  However, 
when all the properties of $\mket{S}$ are applied, we find the surprising result
that {\em all} of the elements of $\mathcal{J}$ are eliminated.
{\em No} assignment of definite results to all six possible measurements
can possibly agree with the correspondences obtained from the entangled
MQT state $\mket{S}$.  We therefore conclude that these correspondences
are incompatible with any local hidden variable theory.

\section*{Mixed states and cloning}

In both actual quantum theory and modal quantum theory,
a {\em mixed state} arises when we cannot ascribe a definite
quantum state vector to a system.  This may happen because
several state vectors are possible, or because the system is
only a part of a larger system in an entangled state.

Suppose that a system in MQT might be in any one of several
possible states $\mket{\psi_{1}}$, $\mket{\psi_{2}}$, etc.
We collect these together into a set $M$, which
characterizes the mixture of states.  An effect is possible
for the mixture $M$ if it is possible for at least one of
the state vectors in $M$.  Equivalently, we say that 
$\mbra{a}$ is {\em impossible} for $M$ provided 
$\mamp{a}{\psi} = 0$ for all $\mket{\psi} \in M$.

Two mixtures $M_{1}$ and $M_{2}$ are equivalent when they
lead to exactly the same possible effects.  Because the effect
functionals are linear, it follows that any mixture $M$
is equivalent to the subspace $\vspan{M}$ spanned by $M$.
Therefore, two mixtures $M_{1}$ and $M_{2}$ will be 
equivalent if $\vspan{M_{1}} = \vspan{M_{2}}$.  If the two
mixtures span different subspaces, then we can always find an
effect (a linear functional) which is possible for one but
not the other.  Therefore, we identify the mixed state 
$\mathcal{M}$ as the subspace $\vspan{M}$ spanned by the
mixture $M$.  In MQT, mixed states are subspaces of $\modalspace$.

How can we arrive at a mixed state for a subsystem of an entangled
system in MQT?  Suppose systems \#1 and \#2 have a joint state
vector $\mket{\psi\sys{12}}$.  Given a basis $\{ \mket{a} \}$
for system \#1, we can write this as
\begin{equation}
    \mket{\psi\sys{12}} = \sum_{a} \mket{a, \psi_{a}} .
\end{equation}
We can take the non-zero states $\mket{\psi_{a}}$ that appear
in this to define a mixture $M$ for system \#2,
which defines in turn a mixed state $\mathcal{M} = \vspan{M}$.
It is straightforward to show that this mixed state for system
\#2 is independent of the choice of basis $\{ \mket{a} \}$ for system \#1.

Finally, we note that a no-cloning theorem holds in MQT, and that
its proof is virtually identical to that of Wootters and Zurek
for AQT \cite{nocloning}.
We imagine a cloning machine that successfully
copies distinct input states $\mket{a}$ and $\mket{b}$, 
a process that can be represented by the evolution of the input, 
output and machine systems:
\begin{equation}
        \mket{a,0,M_{0}} \rightarrow \mket{a,a,M_{a}}
        \quad \mbox{and} \quad
        \mket{b,0,M_{0}} \rightarrow \mket{b,b,M_{b}} .
    \label{cloningsuccess}
\end{equation}
If we now consider the superposition input state
$\mket{c} = \mket{a} + \mket{b}$, linearity of the
overall evolution means that the final state
of input and output is instead either a superposition
or mixture of $\mket{a,a}$ and $\mket{b,b}$ (depending
on the relation of the final machine
states $\mket{M_{a}}$ and $\mket{M_{b}}$).  In neither
case do we obtain the cloned state
$\mket{c,c} = \mket{a,a} + \mket{a,b} + \mket{b,a} + \mket{b,b}$.
Therefore, any cloning machine in MQT must fail for some input
states.

\section*{Superdense coding and teleportation}

One remarkable feature of entangled states in AQT is 
superdense coding \cite{superdense}, 
whereby entanglement can double the
information capacity of a quantum system.  There is a
straightforward analogue of this in MQT.  Consider the
following set of states for two mobits:
\begin{equation}
    \begin{array}{lcl}
        \mket{R} = \mket{0,0} + \mket{1,1} & \quad & 
        \mket{U} = \mket{0,0} + \mket{1,0} + \mket{1,1} \\
        \mket{S} = \mket{0,1} + \mket{1,0} & \quad & 
        \mket{V} = \mket{0,0} + \mket{0,1} + \mket{1,0} 
    \end{array} .
\end{equation}
These four entangled states form a basis, and so may be
identified with the outcome of some measurement.  We also
note that any of the four states can be transformed into 
any other one by invertible linear evolution on one of
the mobits.  Given operators $G$ and $K$ such that
\begin{equation}
    \begin{array}{lcl}
        G \mket{0} = \mket{1} & \quad & K \mket{0} = \mket{0} \\
        G \mket{1} = \mket{0} & \quad & K \mket{1} = \mket{0} + \mket{1}
    \end{array}
\end{equation}
we find that
\begin{equation}
    \mket{S} = G\sys{1} \mket{R} \qquad
    \mket{U} = K\sys{1} \mket{R} \qquad
    \mket{V} = K\sys{1} G\sys{1} \mket{R} .
\end{equation}

Suppose that Alice wishes to send Bob a message by transferring 
a single mobit to him.  She can reliably transmit one bit (two
possible messages), since she can encode the message by two
basis states $\mket{0}$ and $\mket{1}$, which Bob can distinguish
by a $Z$ measurement.  She cannot send more without the possibility
of error; and in any case, there are only three distinct mobit
states available for her to use.

But now suppose instead that Alice and Bob initially share a pair
of mobits in the joint state $\mket{R}$.  Alice can encode two
bits (four possible messages) by choosing to apply the operators
$1$, $G$, $K$ or $KG$ to her mobit, resulting in one of the four
states $\mket{R}$, $\mket{S}$, $\mket{U}$ or $\mket{V}$.  If she
then delivers her transformed mobit to Bob, he can perform a 
joint measurement on both mobits to reliably distinguish these 
possibilities.
This is the MQT analogue of superdense coding.

The same set of entangled mobit states and single-mobit 
transformations can also be used to accomplish the MQT 
analogue of quantum teleportation \cite{teleportation}.

\section*{Remarks}

The mathematical structure of MQT is considerably simpler than
the Hilbert space of AQT.  Without an inner product, MQT
lacks probability amplitudes (and thus probabilities), 
orthogonal bases, and unitary (inner-product-preserving)
evolution.  Without an outer product, MQT cannot represent
mixed states by density operators, or numerical observables
by Hermitian operators.  There is no Hermitian conjugate
($\dagger$) operation.  Furthermore, when $\scalarfield$ is
finite, systems only have a finite number of available states,
and time evolution must be discrete rather than continuous.

Nevertheless, modal quantum theory exemplifies many of the basic
ideas of actual quantum theory, and has analogues for many of
the most striking quantum phenomena.  There is a distinction
between ``classical'' and ``quantum'' modal variables.
Modal quantum systems exhibit superposition and interference
effects, and the time evolution of an isolated system can be 
described by a linear operator.  
These systems display complementarity between incompatible
observables.  The properties of entangled states can be
used to exclude local hidden variable theories and cloning;
they also support information protocols such as superdense 
coding and teleportation.  Finally, mixed states of modal
quantum systems can be naturally identified with the subspaces 
of the modal state space.

Thus, despite its extreme simplicity, MQT is a remarkably 
rich ``toy model'' for quantum physics.

\end{document}